\documentclass[showpacs,preprintnumbers,amsmath,amssymb,superscriptaddress,nofootinbib]{revtex4}

\usepackage{axodraw4j}
\usepackage{pstricks}
\usepackage{color}
\usepackage{graphicx}
\usepackage{dcolumn}
\usepackage{bm}
\usepackage{atlasphysics}

\begin{document}

\title{
Extraction of Top Backgrounds in the Higgs Boson Search with the $H\rightarrow WW^{(*)}\rightarrow\ell\ell+\met$ Decay with a Full-Jet Veto at the LHC}
\author{Bruce Mellado}
\email{bmellado@wisc.edu}
\affiliation{Department of Physics, University of Wisconsin, Madison, WI 53706, USA}
\author{Xifeng Ruan}
\email{ruanxf@mail.ihep.ac.cn}
\affiliation{Institute of High Energy Physics, Chinese Academy of Sciences, Beijing, China}
\affiliation{Laboratoire de l'Acc\'el\'erateur Lin\'eaire, Universit\'e Paris-Sud et IN2P3-CNRS, Orsay, France}
\author{Zhiqing Zhang}
\email{zhangzq@lal.in2p3.fr}
\affiliation{Laboratoire de l'Acc\'el\'erateur Lin\'eaire, Universit\'e Paris-Sud et IN2P3-CNRS, Orsay, France}


\begin{abstract}
\vspace*{0.1cm}
The search for a Standard Model Higgs boson with the $H\rightarrow WW^{(*)}\rightarrow\ell\ell+\met$ decay and the application of a full-jet veto yields a strong sensitivity in the mass range $130<m_H<200\,\gev$. One of the residual backgrounds is related to processes involving top quarks. A method for the extraction of top backgrounds is evaluated at the parton and hadronic levels based on the data-driven extraction of the jet veto survival probability. The dominant systematic uncertainty of the proposed method is estimated to be about 15\%.

\end{abstract}

\pacs{11.15.Ex,14.80.Bn}

\maketitle

\section{Introduction}
\label{sec:intro}

In the Standard Model (SM) of electroweak interactions, there are four types of  gauge vector bosons (gluon, photon, $W^\pm$ and $Z$) and twelve types of fermions (six quarks and six leptons)~\cite{np_22_579,prl_19_1264,sal_1968_bis,Glashow:1970gm}. These particles have been observed experimentally. At present, all the data obtained from the many experiments in particle physics are in agreement with the SM.  In the SM there is one particle, the Higgs boson, that is responsible for generating masses to all of the particles in the theory, the gauge bosons as well as the fermions~\cite{Englert:1964et,pl_12_132,prl_13_508,pr_145_1156,Guralnik:1964eu,pr_155_1554}. In this sense, the Higgs particle occupies a unique position.

The search using the decay mode $H\rightarrow WW^{(*)}\rightarrow\ell\nu\ell^\prime\nu, (\ell=e,\mu)$ gives strong sensitivity to the SM Higgs boson in the intermediate mass range of $130<m_H<200\,\gev$~\cite{PhysRevD.43.779,pr_55_167,PhysRevD.60.113004,Kauer:2000hi,Mellado:2007fb}. The feasibility of Higgs boson searches at the LHC has been extensively studied and reported by the CMS and ATLAS collaborations in Refs.~\cite{CMSPTDR,LHCC99-14,csc,ATL-PHYS-PUB-2010-005} for different proton-proton center-of-mass energies and integrated luminosities scenarios. 

The extraction of the non-resonant $WW$ production and the Higgs boson signal with the $H\rightarrow WW^{(*)}\rightarrow\ell\ell+\met$ decay, where $\met$ is the missing transverse energy from the escaping neutrinos, requires a good understanding of top backgrounds. The region of the phase-space with the best signal-to-background ratio corresponds to a full-jet veto, i.e. that no hadronic jets be observed above a transverse momentum threshold and within a certain pseudorapidity range. These requirements are defined by the characteristics of the experiments and the experimental conditions. The requirement of a full-jet veto is motivated by the need to suppress top backgrounds, the cross-sections of which grow more rapidly with the center-of-mass energy than the Higgs signal processes. Therefore, it is necessary to understand the jet veto survival probability of the top background processes. This is a complex observable. For its determination it is desirable to rely on data or to have the least possible Monte Carlo (MC) dependence. 

In Ref.~\cite{ATL-PHYS-PUB-2010-005} a method for the extraction of the jet veto survival probability ({\sc jvsp}) from data was explored. In that work the quantity {\sc jvsp} was extracted from a control sample made of one high transverse momentum and isolated lepton, large  $\met$ and at least two jets, the invariant mass of which was required to be close to the $W$ mass. Systematic errors of the order of 100\% were reported, though part of the uncertainty is of statistical nature due to the finite size of the MC used for the study. Here we consider the extraction of the jet veto survival probability of top backgrounds from a control sample that comprises two high transverse momentum leptons, \met and a $b$-tagged hadronic jet. This part of phase-space is dominated by top backgrounds. The isolation of top processes experimentally happens in a restricted region of the phase-space. This may introduce a significant bias in the calculation of the jet veto survival probability. This bias and the corresponding theoretical uncertainty are evaluated here. In this method, the full-jet veto survival probability for top backgrounds is expressed as the square of the veto survival probability in events with one $b$-tagged hadronic jet multiplied by a Matrix Element correction.

This paper is organized as follows: Section~\ref{sec:topmc} gives account of the modeling of the top backgrounds; Section~\ref{sec:topevsel} defines the region of the phase-space under study; Section~\ref{sec:topjvsp} discusses the jet veto survival probability at leading order (LO). The impact of QCD higher order corrections and hadronization is discussed in Section~\ref{sec:hoqcdhad}.

\section{Description of Top Processes}
\label{sec:topmc}

As pointed out in Refs.~\cite{pr_62_014021,Kauer:2004fg} the description of top-related processes in the presence of jet vetoes or topological requirements is not trivial. A description of this final state in terms of the sum of the square-amplitudes of $t\overline{t}$ and single top production ($qb\rightarrow tW$ diagrams) is not appropriate for certain parts of the phase-space of interest here. This has been proven to be the case even when finite width effects in the top quark decays are taken into account. Effects have been quantified in the particular context of the Higgs boson search with the $H\rightarrow WW^{(*)}\rightarrow\ell\ell+\met$ decay~\cite{pr_62_014021,Kauer:2004fg}. When calculating the survival probability against a jet veto in a detector it is appropriate to use the full set of diagrams that contribute to $pp\rightarrow W^+W^-b\overline{b}$ processes. Events  with two large transverse momentum leptons and large $\met$ emerging from the $pp\rightarrow W^+W^-b\overline{b}$ processes cannot be split into double resonant, single resonant and non-resonant processes. This problem would be solved with the computation of QCD corrections at next-to-leading-order (NLO) to the complete set of $pp\rightarrow W^+W^-b\overline{b}$ diagrams and their implementation in MC@NLO~\cite{Frixione:2002bd} or POWHEG~\cite{Nason:2004rx,Frixione:2007vw}. Before the appearance of such a tool it is necessary to resort to the LO description. The first results of a NLO QCD calculation for the full set of $pp\rightarrow W^+W^-b\overline{b}$ have been reported recently~\cite{Denner:2010jp}. This represents significant progress towards a proper description of corners of the phase-space of interest here.

The automated program MADGRAPH~\cite{Alwall:2007st} is used to generate MC events corresponding to the various processes considered here. The impact of QCD higher order corrections and hadronization, evaluated with the package MC@NLO~\cite{Frixione:2002bd}, is obtained with the Monte Carlo truth (hadronic)  without the application of detector effects. In both cases default settings are used. The CTEQ6L1 and CTEQ6M~\cite{JHEP_0207_012} parton density parameterizations are used, when appropriate. The renormalization and factorization scales are set to the top mass. No constraint is applied on the transverse momentum of the $b$-partons and these are required to be separated by $\Delta R>0.4$. The decays of the $W$ boson are treated with the narrow width approximation.

The $t\overline{t}$ process is by far the dominant contribution of the top background $pp\rightarrow W^+W^-b\overline{b}$. This is illustrated in Table~\ref{tab:topcrosssections}, where the LO cross-sections in pb calculated using the MADGRAPH package by setting the factorization and renormalization scales to the top mass are shown for the processes $pp\rightarrow W^+W^-b\overline{b}$ and $pp\rightarrow t\overline{t}\rightarrow W^+W^-b\overline{b}$.
The results in Table~\ref{tab:topcrosssections} do not include the branching ratios of the $W$ bosons to leptons.
\begin{table}[htb]
\begin{center}
\begin{tabular}{|c||c|c|}
\hline\hline
Process & 7 \,\tev & 14\,\tev \\ \hline
$pp\rightarrow W^+W^-b\overline{b}$ & 103 & 625 \\
$pp\rightarrow t\overline{t} \rightarrow W^+W^-b\overline{b}$  & 99.6 & 597 \\
\hline\hline
\end{tabular}
\end{center}
\caption{Cross-sections (in pb) at LO for the processes $pp\rightarrow W^+W^-b\overline{b}$ and $pp\rightarrow t\overline{t}\rightarrow W^+W^-b\overline{b}$.
Results are shown for two different values of the center-of-mass energy of the proton-proton collision. The MADGRAPH~\cite{Alwall:2007st} package is used to calculate the cross-sections. \label{tab:topcrosssections}}
\end{table}

\section{Event Selection}
\label{sec:topevsel}

The following requirements are imposed on the final state considered here:
\begin{description}
\item {\bf s1:} two opposite sign leptons (electrons or muons) with $p_T>20\,\gev$ in the pseudorapidity range $\left|\eta\right|<2.5$,
\item {\bf s2:} large missing transverse energy, $\met>30\,\gev$. 
\end{description}

When performing studies at the parton-level the $\met$ is defined as the transverse momentum of neutrinos of the $W$ decay and $b$-tagging is performed by generating a random number. If more than two partons are $b$-tagged the parton that has the largest random number assigned is chosen. The remaining parton is denoted as probe parton. When evaluating the feasibility of top events with a $b$-tagged jet a 60\,\% tagging efficiency is assumed in the range $\left|\eta\right|<2.5$. Experimentally, the $b$-tagging efficiency is a function of the jet transverse momentum. The potential bias introduced by this effect needs to be corrected with MC, as performed in Refs.~\cite{ATLAS-CONF-2011-005,ATLAS-CONF-2011-015}.

The control sample studied here to extract top backgrounds is defined by the additional requirement of a high $p_T$ and central $b$-parton. In addition to the requirements specified above the following condition is imposed:
\begin{description}
\item {\bf s3:} at least one $b$-tagged parton with $p_T>30\,\gev$ in the range  $\left|\eta\right|<2.5$.
\end{description}

\section{Jet Veto Survival Probability and the Definition of the Method}
\label{sec:topjvsp}

The study of the {\sc jvsp} is initially performed at the parton-level using LO matrix elements. The {\sc jvsp} for events with a large $p_T$ gluon in the final state is significantly smaller. The impact of QCD higher order corrections and hadronization will be considered below.  

In this section we define various quantities pertaining to the {\sc jvsp} in events with top backgrounds at LO. It is convenient to define the $P_1$ as the probability of a parton of a particular type to miss a region of phase-space of interest. This region of the phase-space is defined with a lower bound on the transverse momentum, $p_{T}^{v}$, and a maximum value of the parton pseudorapidity, $\eta^{v}$, with the superscript $v$ referring to veto. These are defined by experimental considerations. The hadron-level transverse momentum threshold that is usually applied experimentally is $p_{T}^{v}=20\,\gev$.\footnote{The transverse energy of a reconstructed jet does not in general correspond to an exact value of the parton $p_T$. At LO, requiring no parton with $p_T>30\,\gev$ has a similar efficiency as requiring that no jet with $p_T^v>20\,\gev$ is reconstructed after hadronization in the MC.} In this transverse momentum range the jet finding is not fully efficient.\footnote{In the conditions of ATLAS jet finding becomes fully efficient for $p_T > 40\,\gev$.} The effective value of $p_{T}^{v}$ is expected to be somewhat higher than the nominal value of the threshold that is set experimentally. These two thresholds are related to each other via the experimental inefficiency of observing jets in a range in the vicinity of $p_{T}^{v}$.

The {\sc jvsp} is defined in the control region as the sum of two contributions:
\begin{equation}
P_1 = \left(\int_0^{p_{T}^{v}}\int_{0}^{\left|\eta^{v}\right|} \frac{d\sigma^{\rm cuts}}{dp_{Tb}d|\eta_b|} dp_{Tb}d|\eta_b| + \int_0^{\infty}\int_{|\eta^{v}|}^{\infty} \frac{d\sigma^{\rm cuts}}{dp_{Tb}d|\eta_b|} dp_{Tb}d|\eta_b| \right) / \sigma_{\rm tot}^{\rm cuts},
\label{eq:p1}
\end{equation}
where the integration over the $p_T$ and $|\eta|$ of the $\overline{b}$ parton is assumed and
\begin{equation}
\sigma_{\rm tot}^{\rm cuts} = \int_0^{\infty}\int_0^{\infty} \frac{d\sigma^{\rm cuts}}{dp_{Tb}dp_{T\overline{b}}} dp_{Tb}dp_{T\overline{b}},
\end{equation}
$\sigma^{\rm cuts}$ is the cross-section of the top related backgrounds after the application of the requirements described in Section~\ref{sec:topevsel} where the integration of the pseudorapidity of the partons is implied. The expression for the full-jet veto survival probability at LO can be written as:
\begin{eqnarray}
P_2 &=& \left(\int\int_0^{p_{T}^{v}}\int\int_{0}^{\left|\eta^{v}\right|} \frac{d\sigma^{\rm cuts}}{dp_{Tb}dp_{T\overline{b}}d|\eta_b|d|\eta_{\overline{b}}|} dp_{Tb}dp_{T\overline{b}}d|\eta_b|d|\eta_{\overline{b}}| \right. \nonumber \\
&+& 2\int_0^{\infty}\int_{0}^{\left|\eta^{v}\right|}\int_0^{p_{T}^{v}}\int_{|\eta^{v}|}^{\infty} \frac{d\sigma^{\rm cuts}}{dp_{Tb}dp_{T\overline{b}}d|\eta_b|d|\eta_{\overline{b}}|} dp_{Tb}d|\eta_b| dp_{T\overline{b}}d|\eta_{\overline{b}}| \nonumber \\
&+&
\left. \int_0^{\infty}\int_{|\eta^{v}|}^{\infty}\int_0^{\infty}\int_{|\eta^{v}|}^{\infty} \frac{d\sigma^{\rm cuts}}{dp_{Tb}dp_{T\overline{b}}d|\eta_b|d|\eta_{\overline{b}}|} dp_{Tb}d|\eta_b| dp_{T\overline{b}}d|\eta_{\overline{b}}| \right) /  \sigma_{\rm tot}^{\rm cuts}\,.
\label{eq:p2}
\end{eqnarray}
The calculation of $P_1$ and $P_2$ is performed after the application of requirements {\bf s1} and {\bf s2} (see Section~\ref{sec:topevsel}). It is important to note that the second term in Expression~(\ref{eq:p1}) and the last two terms in  Expression~(\ref{eq:p2}) become significantly smaller than the first terms for values of $|\eta^{v}|>3$. These terms become negligible for $|\eta^{v}|>4$.

The survival probability relevant to the control sample used here is defined as:
\begin{equation}
P_1^{\rm Btag} = \left(\int_0^{p_{T}^{v}}\int_{0}^{\left|\eta^{v}\right|} \frac{d\sigma^{\rm Btag}}{dp_{T}d|\eta|} dp_{T}d|\eta|  + \int_0^{\infty}\int_{\left|\eta^{v}\right|}^\infty \frac{d\sigma^{\rm Btag}}{dp_{T}d|\eta|} dp_{T}d|\eta| \right) / \sigma^{\rm Btag}_{\rm tot},
\label{eq:p1btag}
\end{equation}
where $\frac{d\sigma^{\rm Btag}}{dp_{T}d|\eta|}$ and $\sigma^{\rm Btag}_{\rm tot}$ are the double differential and the total cross-section  after the application of the requirements described in Section~\ref{sec:topevsel}, including the requirement that the event has at least one $b$-tagged parton (requirements {\bf s1}-{\bf s3} in Section~\ref{sec:topevsel}). It is worth noting that the second term in Expression~(\ref{eq:p1btag}) is significantly smaller than the first.

The probability $P_1^{\rm Btag, Exp}$ experimentally can be obtained from the reconstructed probe jet multiplicity distribution of the $b$-tagged control sample by
\begin{equation}
P_1^{\rm Btag, Exp}=\frac{N_{\rm Top}^{\rm Exp}(\ell\ell+\met+b\mbox{-tagged}, 0j)}{N_{\rm Top}^{\rm Exp}(\ell\ell+\met+b\mbox{-tagged})}\label{eq:p1mc}
\end{equation}
where the numerator refers to the number of events having zero probe jet (excluding the $b$-tagged jet in the event) and the denominator is the total number of events of the $b$-tagged samples. The corresponding experimental full-jet veto survival probability $P_2^{\rm Exp}$ can be derived using the jet multiplicity distribution:
\begin{equation}
P_2^{\rm Exp}=\frac{N_{\rm Top}^{\rm Exp}(\ell\ell+\met, 0j)}{N_{\rm Top}^{\rm Exp}(\ell\ell+\met)}\label{eq:p2mc}
\end{equation}
where the numerator and the denominator are the number of top background events after the application of a full-jet veto and the total number of top background events without the application of any requirements on the jet multiplicity in the final state, respectively.

When these probabilities are much smaller than one for low $p_T$ thresholds, one can assume $P_2^{\rm Exp}\approx \epsilon_0^2P_2$ and $P_1^{\rm Btag, Exp}\approx \epsilon_0 P_1^{\rm Btag}$ where $\epsilon_0$ is an experimental efficiency of observing a low $p_T$ jet in the region of interest. These relations may be rewritten as
\begin{equation}
\frac{P_2^{\rm Exp}}{\left(P_1^{\rm Btag, Exp}\right)^2}\approx \frac{P_2}{\left(P_1^{\rm Btag}\right)^2}\,.
\label{eq:p2mcratio}
\end{equation}
Here the experimental uncertainties related to the hadronic energy scale and jet finding inefficiencies would cancel out.

The data-driven prediction will rely on the rate of top-related events observed in data:
\begin{equation}
N^{\rm Exp}_{\rm Top}(\ell\ell+\met,0j) 
\approx N^{\rm Exp}_{\rm Top}(\ell\ell+\met)\times P_2^{\rm Exp} \approx N_{\rm Top}^{\rm Exp}(\ell\ell+\met)\left(P_1^{\rm Btag, Exp}\right)^2  \frac{P_2}{\left(P_1^{\rm Btag}\right)^2}\,.
\label{eq:topmaster}
\end{equation}
The rejection of processes not related to top backgrounds for the measurement of $N^{\rm Exp}_{\rm Top}(\ell\ell+\met)$ is implied. The production of two isolated leptons and large $\met$ at center-of-mass energies studied in this paper is dominated by top-related processes. The contribution of non-top related backgrounds is less than 30\% which is dominated by the SM $WW$ process~\cite{ATLAS-CONF-2011-005,ATLAS-CONF-2011-111}. The systematics related to the extraction of these backgrounds are estimated to be $<3\%$ and thus small compared to other uncertainties considered here. In this master formula uncertainties related to the luminosity measurement and the residual theoretical uncertainties of the total cross-section of top processes cancel out. Ultimately, the data-driven prediction will rely on the correction factor $P_2/(P_1^{\rm Btag})^2$ computed by a MC, $P_2^{\rm MC}/(P_1^{\rm Btag,MC})^2$. In this paper we focus on the relevant properties of this correction factor computed at parton level and at LO. The impact of hadronization and higher order corrections are discussed in Section~\ref{sec:hoqcdhad}.

Table~\ref{tab:jvsp7tevtt} displays the results of jet veto survival probabilities for $pp\rightarrow W^+W^-b\overline{b}$ processes (see Section~\ref{sec:topmc}) at 7\,\tev center-of-mass energy. Results are shown for different values of the transverse momentum for the jet veto. The second, third columns show the values of $P_1$ and $P_2$, respectively. The fourth column displays the ratio $P_1^2/P_2$. Although $P_1$ does not necessarily lead to a useful observable, the ratio $P_1^2/P_2$ helps us understand the potential angular correlations between the $b$-partons. A deviation from unity is indicative of a correlation. This seems to be the case for low values of $p_{T}^{v}$ and the effect yields $\approx 7\%$ correction for $p_{T}^{v}=30\,\gev$. This correlation appears due to kinematics, in events where the products of the reaction go in the forward region (asymmetric collisions). The fifth and sixth columns show the value of $P_1^{\rm Btag}$ and $P_1^{\rm Btag}/P_1$, respectively. The latter quantifies the bias introduced by determining the {\sc jvsp} from a part of the phase-space where a $b$-parton is required. The bias seems moderate and constitutes 25\% for $p_{T}^{v}=30\,\gev$. The last column displays the final correction factor $(P_1^{\rm Btag})^2/P_2$ that will be used in Expression~(\ref{eq:topmaster}). 
\begin{table}[t]
\begin{center}
\begin{tabular}{|c||c|c|c||c|c|c|}
\hline\hline
$p_{T}^{v} [\gev]$ & $P_1$ & $P_2$ & $P_1^2/P_2$ & $P_1^{\rm Btag}$ & $P_1^{\rm Btag}/P_1$ & $(P_1^{\rm Btag})^2/P_2$  \\ \hline
  20.0 &   0.07 &  0.005 &   0.89 &   0.09 &   1.29 &   1.50  \\ 
  22.5 &   0.08 &  0.007 &   0.89 &   0.10 &   1.27 &   1.45  \\ 
  25.0 &   0.10 &  0.010 &   0.94 &   0.12 &   1.27 &   1.51  \\ 
  27.5 &   0.11 &  0.013 &   0.95 &   0.14 &   1.27 &   1.53  \\ 
  30.0 &   0.13 &  0.018 &   0.93 &   0.16 &   1.25 &   1.46  \\ 
  32.5 &   0.15 &  0.025 &   0.90 &   0.18 &   1.21 &   1.32  \\ 
  35.0 &   0.17 &  0.032 &   0.94 &   0.21 &   1.18 &   1.31  \\ 
  37.5 &   0.20 &  0.040 &   0.99 &   0.23 &   1.16 &   1.32  \\ 
  40.0 &   0.23 &  0.051 &   1.01 &   0.26 &   1.14 &   1.30  \\ 
\hline\hline
\end{tabular}
\end{center}
\caption{Results of jet veto survival probabilities for $pp\rightarrow W^+W^-b\overline{b}$ processes (see Section~\ref{sec:topmc}) at 7\,\tev center-of-mass energy. Results are shown for different values of the transverse momentum for the jet veto.  \label{tab:jvsp7tevtt}}
\end{table}

It is relevant to note that the dependence of $(P_1^{\rm Btag})^2/P_2$ on the transverse momentum threshold is mild. The results are stable within better than 15\% over a large $p_T^v$ range between $20$ and $40$\,GeV. This effectively leads us to conclude that the effect of hadronic energy scale uncertainties will have little impact on $(P_1^{\rm Btag})^2/P_2$. 

Table~\ref{tab:jvsp7tevttonly} shows the same results as in Table~\ref{tab:jvsp7tevtt} but for $t\overline{t}$ production only. The values of the {\sc jvsp} in Table~\ref{tab:jvsp7tevttonly} are significantly lower than for the full $pp\rightarrow W^+W^-b\overline{b}$ treatment. This is expected. What is particularly relevant for the robustness of the method is the fact that the ratio $(P_1^{\rm Btag})^2/P_2$ is stable well within typically $10\%$.  The ratio $P_1^2/P_2$ shows a smaller bias with respect to that observed for the full $pp\rightarrow W^+W^-b\overline{b}$ processes. 
\begin{table}[htb]
\begin{center}
\begin{tabular}{|c||c|c|c||c|c|c|}
\hline\hline
$p_{T}^{v} [\gev]$ & $P_1$ & $P_2$ & $P_1^2/P_2$ & $P_1^{\rm Btag}$ & $P_1^{\rm Btag}/P_1$ & $(P_1^{\rm Btag})^2/P_2$  \\ \hline
  20.0 &   0.06 &  0.003 &   1.05 &   0.07 &   1.19 &   1.49  \\ 
  22.5 &   0.07 &  0.004 &   1.11 &   0.08 &   1.19 &   1.58  \\ 
  25.0 &   0.08 &  0.007 &   1.03 &   0.10 &   1.19 &   1.47  \\ 
  27.5 &   0.10 &  0.010 &   0.97 &   0.12 &   1.19 &   1.36  \\ 
  30.0 &   0.12 &  0.014 &   1.01 &   0.14 &   1.19 &   1.44  \\ 
  32.5 &   0.14 &  0.019 &   1.01 &   0.16 &   1.17 &   1.38  \\ 
  35.0 &   0.16 &  0.026 &   0.99 &   0.18 &   1.14 &   1.28  \\ 
  37.5 &   0.18 &  0.034 &   0.99 &   0.21 &   1.12 &   1.24  \\ 
  40.0 &   0.21 &  0.042 &   1.03 &   0.23 &   1.09 &   1.23  \\ 
\hline\hline
\end{tabular}
\end{center}
\caption{Same as Table~\ref{tab:jvsp7tevtt} for $t\overline{t}$ production only.  \label{tab:jvsp7tevttonly}}
\end{table}

Table~\ref{tab:jvsp14tev} shows the results for the $pp\rightarrow W^+W^-b\overline{b}$ at 14\,\tev center-of-mass energy. The results for $(P_1^{\rm Btag})^2/P_2$ display a mild increase with respect to the results obtained for 7\,\tev. 
\begin{table}[t]
\begin{center}
\begin{tabular}{|c||c|c|c||c|c|c|}
\hline\hline
$p_{T}^{v} [\gev]$ & $P_1$ & $P_2$ & $P_1^2/P_2$ & $P_1^{\rm Btag}$ & $P_1^{\rm Btag}/P_1$ & $(P_1^{\rm Btag})^2/P_2$  \\ \hline
  20.0 &   0.07 &  0.005 &   1.12 &   0.09 &   1.23 &   1.71  \\ 
  22.5 &   0.09 &  0.007 &   1.04 &   0.10 &   1.20 &   1.51  \\ 
  25.0 &   0.10 &  0.010 &   1.01 &   0.12 &   1.22 &   1.50  \\ 
  27.5 &   0.12 &  0.013 &   1.07 &   0.14 &   1.21 &   1.55  \\ 
  30.0 &   0.14 &  0.018 &   1.05 &   0.17 &   1.22 &   1.55  \\ 
  32.5 &   0.15 &  0.022 &   1.08 &   0.18 &   1.19 &   1.51  \\ 
  35.0 &   0.18 &  0.030 &   1.04 &   0.20 &   1.15 &   1.37  \\ 
  37.5 &   0.20 &  0.038 &   1.04 &   0.22 &   1.14 &   1.34  \\ 
  40.0 &   0.22 &  0.047 &   1.04 &   0.25 &   1.11 &   1.28  \\ 
\hline\hline
\end{tabular}
\end{center}
\caption{Results of jet veto survival probabilities for $pp\rightarrow W^+W^-b\overline{b}$ at 14\,\tev center-of-mass energy. Results are shown for different values of the transverse momentum for the jet veto.  \label{tab:jvsp14tev}}
\end{table}

Because the transverse momentum spectrum of partons is relatively steep it is relevant to check the impact of the detector resolution. The energy of partons is smeared according to a single Gaussian distribution to mimic a resolution with a stochastic term of $0.75/\sqrt{E}$. The results for $(P_1^{\rm Btag})^2/P_2$ are stable within a few percent.

As seen in Table~\ref{tab:jvsp7tevtt} the kinematic bias introduced by requiring  $b$-tagging in a restricted region of the phase-space is not trivial and needs to be corrected in Expression~(\ref{eq:topmaster}). In order to evaluate the stability of the results against changes in the transverse momentum of $b$-tagging, results are obtained for $p_T^{v}=20,40\,\gev$. These variations are significantly larger than the energy scale uncertainties for low $p_T$ jets. The results for $(P_1^{\rm Btag})^2/P_2$ and $p_{T}^{v}=30\,\gev$ are stable within 15\%. This estimate is the maximum deviation observed in $(P_1^{\rm Btag})^2/P_2$ in a relatively wide range of the jet veto. In order to check the stability of the results for the range in $\left|\eta\right|$ chosen for the $b$-tagging the range is changed to $\left|\eta\right|<2$. The results for $(P_1^{\rm Btag})^2/P_2$ and $p_{T}^{v}=30\,\gev$ are stable within 10\,\%. Finally, the pseudorapidity range of the jet veto is changed to $\left|\eta\right|<4$. The results for $(P_1^{\rm Btag})^2/P_2$ and $p_{T}^{v}>30\,\gev$ show little change.

Variations of the factorization and renormalization scales yield changes of $(P_1^{\rm Btag})^2/P_2$ that are consistent with the statistics of the samples used. These are much smaller than 15\%.

From these studies it is concluded that the results for $(P_1^{\rm Btag})^2/P_2$ at $p_{T}^{v}=30\,\gev$ are stable within 15\% for the region of the phase-space discussed here. Variations in the QCD scales results in much smaller variations of the results for $(P_1^{\rm Btag})^2/P_2$. These variations are consistent with the statistics of the MC samples generated for the studies reported here.

\section{Impact of QCD Higher Order Corrections, Hadronization and Pile-up}
\label{sec:hoqcdhad}

The ratio $(P_1^{\rm Btag})^2/P_2$ may be affected by a number of experimental effects that can be studied in detail with a full-detector MC simulation, even though their impact has been shown to be small~\cite{ATLAS-CONF-2011-005}. This pertains to the transverse momentum dependence of $b$-tagging, pseudorapidity dependence of the hadronic energy scale, the impact of pile-up, etc...  As a result, experimentalists will replace the correction $P_2/(P_1^{\rm Btag})^2$ in Expression~(\ref{eq:topmaster}) by $P_2^{\rm MC}/(P_1^{\rm Btag,MC})^2$ leading to 
\begin{equation}
N^{\rm Exp}_{\rm Top}(\ell\ell+\met,0j) 
\approx  N_{\rm Top}^{\rm Exp}(\ell\ell+\met)\left(P_1^{\rm Btag, Exp}\right)^2  \frac{P_2^{\rm MC}}{\left(P_1^{\rm Btag,MC}\right)^2}\,.
\label{eq:topmasterMC}
\end{equation}

Because the results for $(P_1^{\rm Btag})^2/P_2$ vary little with respect to that obtained with $pp\rightarrow t\overline{t}\rightarrow W^+W^-b\overline{b}$ the impact of QCD higher order corrections and hadronization are checked here with a large sample of $t\overline{t}$ events produced with MC@NLO. The {\sc jvsp} defined in Section~\ref{sec:topjvsp} were obtained for a transverse momenta thresholds of $p_T=20\,\gev$ (both for the jet veto and the $b$-tagging). 
In order to minimize the effect of final state radiation of gluons off the $b$-parton it is required that the {\sc jvsp} is calculated with jets that pass the requirement of $\Delta R>1$ with respect to the tagging $b$-parton. In this setup $(P_1^{\rm Btag})^2/P_2$, calculated at hadron level, decreases by approximately 10\,\%. These results have been confirmed with large samples that include the full simulation of the ATLAS detector~\cite{ATLAS-CONF-2011-005,ATLAS-CONF-2011-111}.

The effect of additional soft proton-proton collisions, or pile-up has been studied in the past for different scenarios~\cite{CMSPTDR,LHCC99-14,csc}. The impact of pile-up can be properly taken into account in the calculation of the ratio $(P_1^{\rm MC})^2/P_2^{\rm MC}$ with MC samples which describe the real data, e.g.\ in terms of the number of primary vertexes distribution. This implies that the MC needs to have the correct simulation of pile-up effects. Provided that the transverse momentum of the jet veto is large enough pile-up effects have been shown to be small~\cite{ATLAS-CONF-2011-111}.

\section{Conclusions and Prospects}

Top processes are an  important background for Higgs boson search with the $H\rightarrow WW^{(*)}\rightarrow\ell\ell+\met$ decay. In order to isolate the Higgs boson search a full-jet veto needs to be applied. In this paper a method for the extraction of the jet veto survival probability of top backgrounds is proposed. The full-jet veto survival probability for top backgrounds is expressed as the square of the veto survival probability in events with one $b$-tagged hadronic jet multiplied by a Matrix Element correction, $(P_1^{\rm Btag})^2/P_2$. The values of $(P_1^{\rm Btag})^2/P_2$ vary little when going from the complete set of Matrix Elements $pp\rightarrow W^+W^-b\overline{b}$ to $pp\rightarrow t\overline{t}\rightarrow W^+W^-b\overline{b}$. Therefore, in practice this Matrix Element correction can be obtained with a $t\overline{t}$ Monte Carlo.

The error of the method is dominated by the lack of knowledge of the parton transverse momenta, leading to an uncertainty of 15\% in the ratio $(P_1^{\rm Btag})^2/P_2$. The variation due to changes in the factorization and normalization scales yields a significantly smaller effect. A rigorous evaluation of the impact of QCD higher order corrections on $(P_1^{\rm Btag})^2/P_2$ would require a complete calculation for the complete set of $pp\rightarrow W^+W^-b\overline{b}$ diagrams. This exercise will be possible in the near future since the first results of a NLO QCD calculation for the full set of $pp\rightarrow W^+W^-b\overline{b}$ are already available~\cite{Denner:2010jp}.

\begin{acknowledgments} 
The authors would like to thank N.~Kauer and Y.~Pan for discussions.
This work was supported in part by the DoE Grant No. DE-FG02-95ER40896. The work of B.M. is also supported by awards from  the Wisconsin Alumni Research Foundation and the Vilas Foundation.
\end{acknowledgments}

\bibliography{vbf,mycites,atlaspaper}
\end{document}